%% file: atomgnn.tex
\documentclass[fleqn,10pt]{wlscirep}
\usepackage[utf8]{inputenc}
\usepackage[T1]{fontenc}
\usepackage[square,sort,comma,numbers]{natbib}

\usepackage{helvet} % DO NOT CHANGE THIS
\usepackage{algorithm}
\usepackage{algpseudocode}
\usepackage{booktabs}

\usepackage{courier}  % DO NOT CHANGE THIS
\usepackage{graphicx} % DO NOT CHANGE THIS
\usepackage{bm}
\usepackage{amsthm}
\usepackage[switch]{lineno}  %

\usepackage{natbib}  % DO NOT CHANGE THIS AND DO NOT ADD ANY OPTIONS TO IT
\usepackage{caption} % DO NOT CHANGE THIS AND DO NOT ADD ANY OPTIONS TO IT
\usepackage{amsfonts}
\usepackage{booktabs}
\usepackage{multirow}
\usepackage{bm}
\usepackage{amsmath}

\input{section/commend}
\title{Distance-aware Molecule Graph Attention Network for Drug-Target Binding
Affinity Prediction}

\author[1]{Jingbo Zhou}
\author[1,2]{Shuangli Li}
\author[1]{Liang Huang}
\author[1]{Haoyi Xiong}
\author[1]{Fan Wang}
\author[2]{Tong Xu}
\author[3]{Hui Xiong}
\author[1]{Dejing Dou}

\affil[1]{Baidu Inc., Email: \{zhoujingbo, lishuangli, lianghuang,xionghaoyi,wangfan04,doudejing\}@baidu.com.}
\affil[2]{University of Science and Technology of China, Email: tongxu@ustc.edu.cn.}
\affil[3]{Rutgers, The State University of New Jersey, Email:hxiong@rutgers.edu.}

\input{section/sec0-abstract}

\begin{document}
%\linenumbers
\maketitle

\input{section/sec1-introduction}
\input{section/sec2-related}
\input{section/sec3-preliminaries}
\input{section/sec4-method}
\input{section/sec5-experiment.tex}
\input{section/sec6-conclusion}

% \section{References}

% \section{ Acknowledgments}

%\bibliographystyle{aaai21}
\bibliography{atomgnn.bib}
\end{document}

%% file: section/commend.tex
\usepackage{amsmath}
\usepackage{amssymb}
\usepackage{color}
\usepackage{subfigure}

\usepackage{graphicx}
\usepackage{csquotes}
\usepackage{url}
\usepackage{epstopdf}
\usepackage{algorithm} 
\makeatletter
\newif\if@restonecol
\makeatother

\usepackage[ruled,vlined,boxed,linesnumbered,algo2e]{algorithm2e}
\newcommand{\hide}[1]{} % use this instead of comment environment

\usepackage{multirow}
\usepackage{siunitx}

\usepackage[normalem]{ulem}
\usepackage{soul}
\usepackage{xcolor}

\newcommand{\hidespace}[1]{{\hide{#1}}} 

\newcommand{\B}[1]{{\bfseries #1}}
\newcommand{\model}{S-MAN\xspace}
\newcommand{\gnn}{S-MAN\xspace}
\newcommand{\graph}{SPoG\xspace} % Spatial-enhanced Pocket-ligand Graph SPoG

%% file: section/sec0-abstract.tex
\begin{abstract}

Accurately predicting the binding affinity between drugs and proteins is an essential step for computational drug discovery.  
Since graph neural networks (GNNs) have demonstrated remarkable success in various graph-related tasks, GNNs have been considered as a promising tool to improve the binding affinity prediction in recent years. However, most of the existing GNN architectures can only encode the topological graph structure of drugs and proteins without considering the relative spatial information among their atoms. Whereas, different from other graph datasets such as social networks and commonsense knowledge graphs, the relative spatial position and chemical bonds among atoms have significant impacts on the binding affinity. To this end, in this paper, we propose a diStance-aware Molecule graph Attention Network (S-MAN) tailored to drug-target binding affinity prediction. As a dedicated solution, we first propose a position encoding mechanism to integrate the topological structure and spatial position information into the constructed pocket-ligand graph. Moreover, we propose a novel edge-node hierarchical attentive aggregation structure which has edge-level aggregation and node-level aggregation. The hierarchical attentive aggregation can capture spatial dependencies among atoms, as well as fuse the position-enhanced information with the capability of discriminating  multiple spatial relations among atoms. Finally, we conduct extensive experiments on two standard datasets to demonstrate the effectiveness of \model.

\end{abstract}

%% file: section/sec1-introduction.tex
\section{Introduction}

% Spatial-enhanced Pocket-ligand Graph (SPoG)
% Position-aware Molecule Graph Attention Network (PMAN)
% Drug-target binding affinity prediction (DTA)
% Drug-target interaction prediction (DTI)
% drugprotein interactions (DPIs)
% Drug / ligand, target / protein

% 1.introduce the drug and target, 2.describe the binding affinity, the importance and the traditional methods, 3.some related work about DTA and limitation, 4.our work
Drug-target binding affinity (DTA) prediction has been widely considered as one of the most important tasks in computational drug discovery for a long time \cite{drews2000drug}.
%Drug-target binding affinity (DTA) prediction is one of the important tasks in drug discovery \cite{drews2000drug}. 
Drugs are chemical compounds, which can be represented by a molecular graph in general. Drugs and small molecules are also called ligands which can react with targets. Usually, targets are referred to as commonly proteins, such as enzymes, ion channels and receptors, which can activate or inhibit a biological process to cure a disease after binding with a ligand.
The quantity of binding strength (measured by a real number) among the drug-target interaction is referred to as \textit{binding affinity} which is an important concept related to the treatment of diseases. 

\begin{figure}[t]
\centering
\includegraphics[width=0.7\columnwidth]{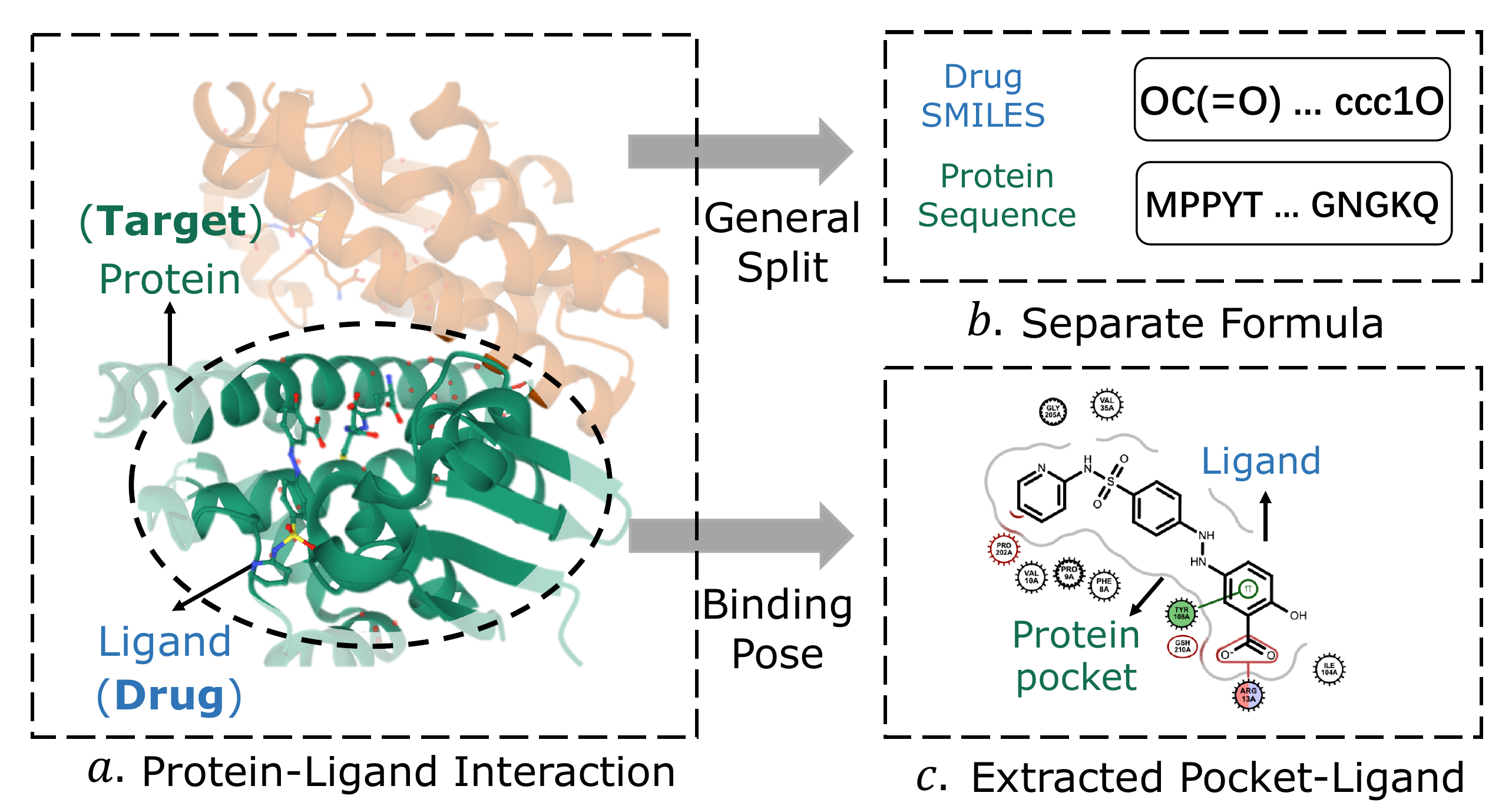}
%\vspace{-5mm}
% \caption{An illustrative example of visualizing protein-ligand interactions (PDB: 13GS).}
\caption{An illustrative example (PDB: 13GS) of converting the protein-ligand interactions (in figure $a$) to the input of model in two different ways (in figure $b$ and $c$).}
\label{lig-pock}
\vspace{-5mm}
\end{figure}

\B{Example 1.} \textit{ Taking Figure \ref{lig-pock} for example, there are two main objects in DTA prediction: 1) A drug (or ligand) can interact with the 2) target (or protein) at the specific circular area as shown in Figure \ref{lig-pock}($a$). DTA prediction aims to output the strength of the interactions of the pairs. One way to do that is to separate the protein-ligand complexes into two sequence formulas as shown in Figure \ref{lig-pock}($b$) and then input them into a prediction function; while Figure \ref{lig-pock}($c$) demonstrates another way to extract the specific binding pose (i.e., pocket-ligand) for DTA prediction.}

Conventionally, the drug-target binding affinity can be estimated by high-throughput screening experiments, which is a costly and time-consuming process \cite{cohen2002protein,noble2004protein}. Predicting DTA can help to accelerate the virtual screening of compounds, which reduces the time and cost of high-throughput screening data by cherry-picking compounds \cite{butler2018machine,ekins2019exploiting,mcgaughey2007comparison}. Therefore, accurately and effectively predicting DTA which can bring great economical benefits for developing new drugs has attracted significant research attention in past decades \cite{wang2003comparative,ballester2010machine,emig2013drug, wen2017deep, zheng2020predicting}.

%A drug can react with the protein, and the quantity of binding for the treatment of diseases is referred to \textit{binding affinity}. 
%Accurately and effectively predicting DTA will bring great economical benefit and help to develop new drugs in the medical field, which can contribute to finding out more valuable candidate drugs.

%Traditionally, experimental assay is the surest way to obtain the binding affinity, while it's expensive and time-consuming to analyze plenty of possible drug-target pairs. As a result, DTA prediction has attracted considerable attention in recent years \cite{emig2013drug, wen2017deep, zheng2020predicting}. 

The early studies of DTA mainly focus on developing physic-based methods and machine-learning methods. Specific domain knowledge is required to design scoring functions \cite{wang2003comparative} and extract features for physic-based methods. Machine learning methods also utilize the well-designed features \cite{ballester2010machine} for DTA prediction. However, these methods relying on feature engineering and useful rules suffer from the problem of limited accuracy and generality on large datasets.

With the development of deep learning,
%\li{some works based on convolutional neural network (CNN)  are also exploited for DTA prediction. } 
convolutional neural networks (CNNs)  are also exploited in some studies for DTA prediction.
Both 1D-CNN model \cite{ozturk2018deepdta} applied to drug-protein sequence and 3D-CNN model \cite{stepniewska2018development} treating drug-target complexes as 3D images are investigated. 
%spatial structures were proven effective. 
In fact, both drugs (a.k.a ligands) and targets (a.k.a proteins) are 3D graphs, which preserve the position information of atoms and relations of covalent bonds. \hidespace{All the}CNN-based methods cannot directly encode such structure information of compounds, and these machine learning models can only be trained in an end-to-end fashion with requiring a large dataset. However, because the high-throughput screening  experiments are high-cost and time-consuming, the size of the available DTA dataset is always small, which degrades the performance of CNN methods. %Moreover, treating the drug-target complexes as 3D images might also cause the inconsistency of spatial information of drugs in different complexes.

%With the development of deep learning and graph representation learning, some works based on convolutional neural network (CNN) and graph neural network (GNN) are also exploited for DTA prediction. Both 1D-CNN model \cite{ozturk2018deepdta} applied to drug and protein sequence and 3D-CNN model applied to spatial structures were proven effective. However, treating the drug-target complexes as 3D images might cause the inconsistency of spatial information in different complexes. 

\begin{figure}
  \centering
  \subfigure[]{
    \label{sp-max-dist} %% label for second subfigure
    \includegraphics[width=0.3\columnwidth]{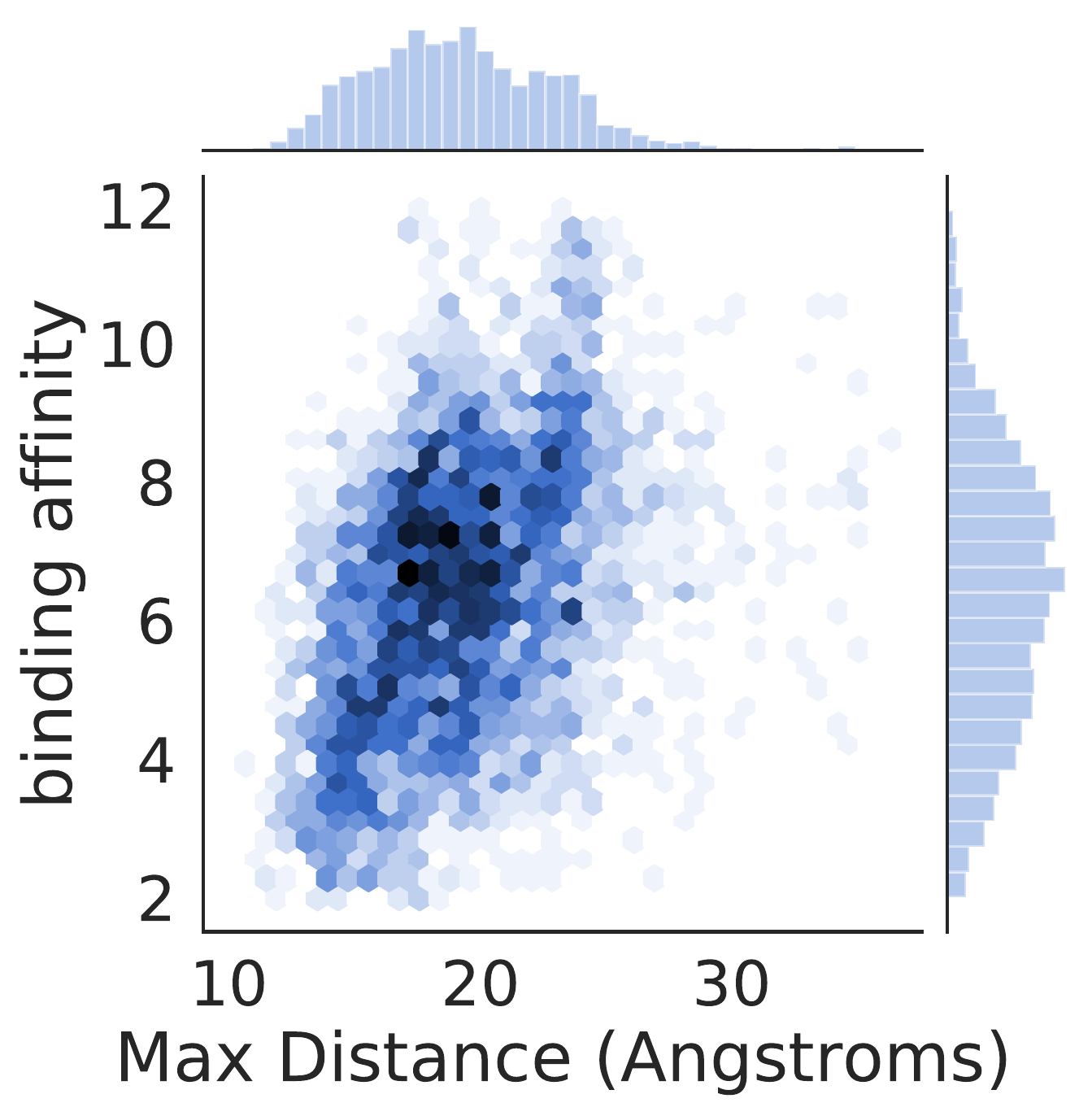}}
      \subfigure[]{
    \label{sp-distribution} %% label for first subfigure
    \includegraphics[width=0.3\columnwidth]{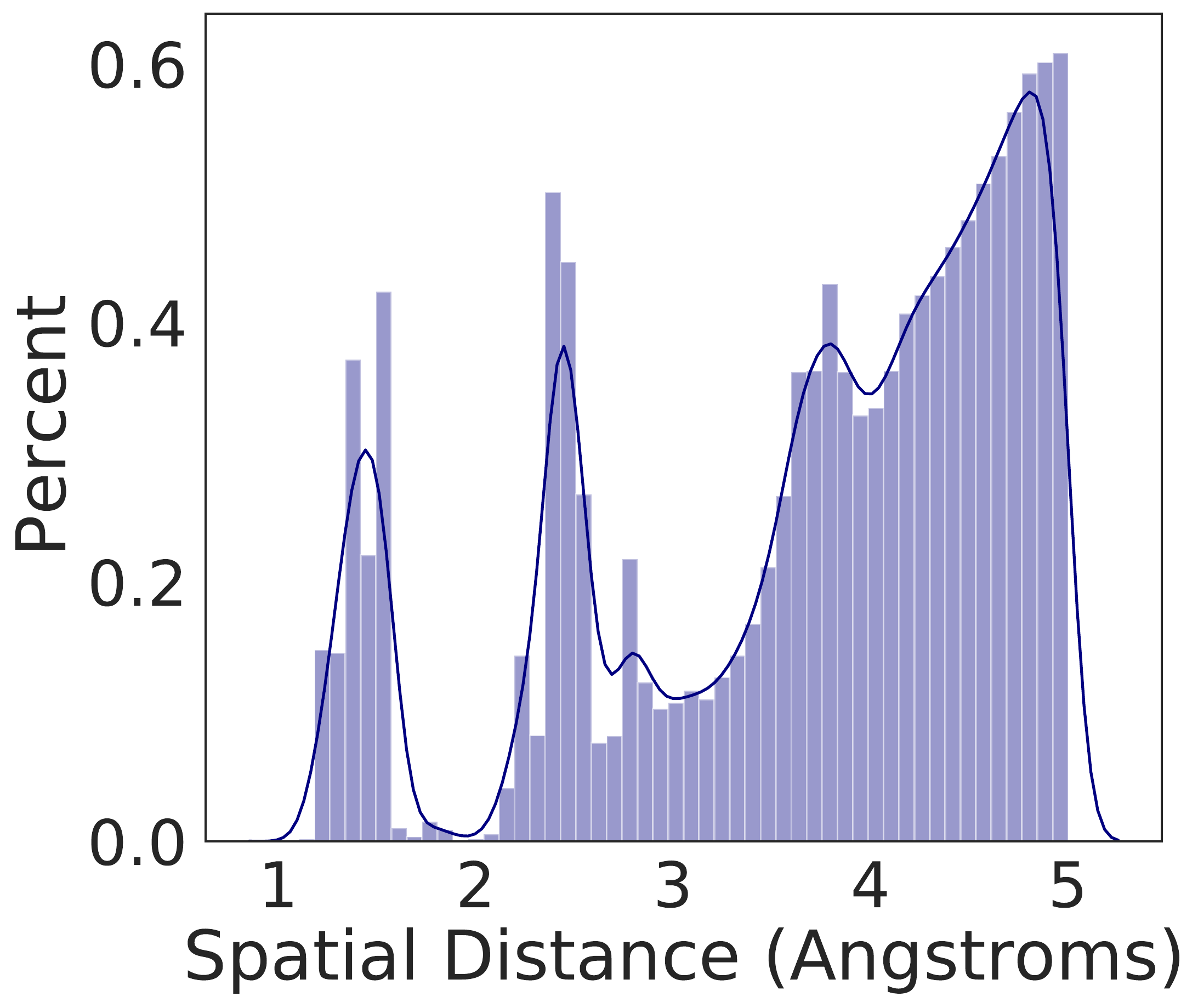}}
  \vspace{-3mm}
  \caption{Spatial distance visual analysis. (a) The correlation between the bindding affinity and the max distance of atoms in  drug-target complex; (b) The distribution of spatial distance between atoms within 5 \si{\angstrom} in  drug-target complex.}
  \vspace{-3mm}
  \label{sp-dist} %% label for entire figure
\end{figure}

More recently, graph neural network (GNN) models have exhibited \hidespace{their} powerful ability of learning molecular graph for DTA prediction \cite{nguyen2020graphdta,lin2020deepgs}, since they can incorporate the graph structure indicated by covalent bonds into the models.  Although these models have achieved great performance, there are two obvious drawbacks: (I) Most existing GNN models can only treat the compounds as a topological graph whose nodes are atoms and whose edges are covalent bonds. The spatial information of relative position among atoms in drug-target complexes is not taken into consideration. However, \hidespace{the} binding affinity is related closely to such spatial position information. An evidence as illustrated in Figure \ref{sp-max-dist} is that the binding affinity has a strong correlation to the max distance 
among atoms of drug-target complexes. (II) Current approaches mostly follow the framework of question-answering-like Siamese network styles to learn the drug and protein representations separately \cite{gao2018interpretable} as shown in Figure \ref{lig-pock}(b). %\li{For example, a GNN module is employed to learn the graph-level embedding of drug, and the protein features can be extracted by a 1D-CNN or RNN module} \cite{gao2018interpretable}. 
Such a separate learning structure leads to the insufficiency of modeling the interactions between drug and protein without considering spatial information.

\begin{figure}
\centering
\includegraphics[width=0.45\columnwidth]{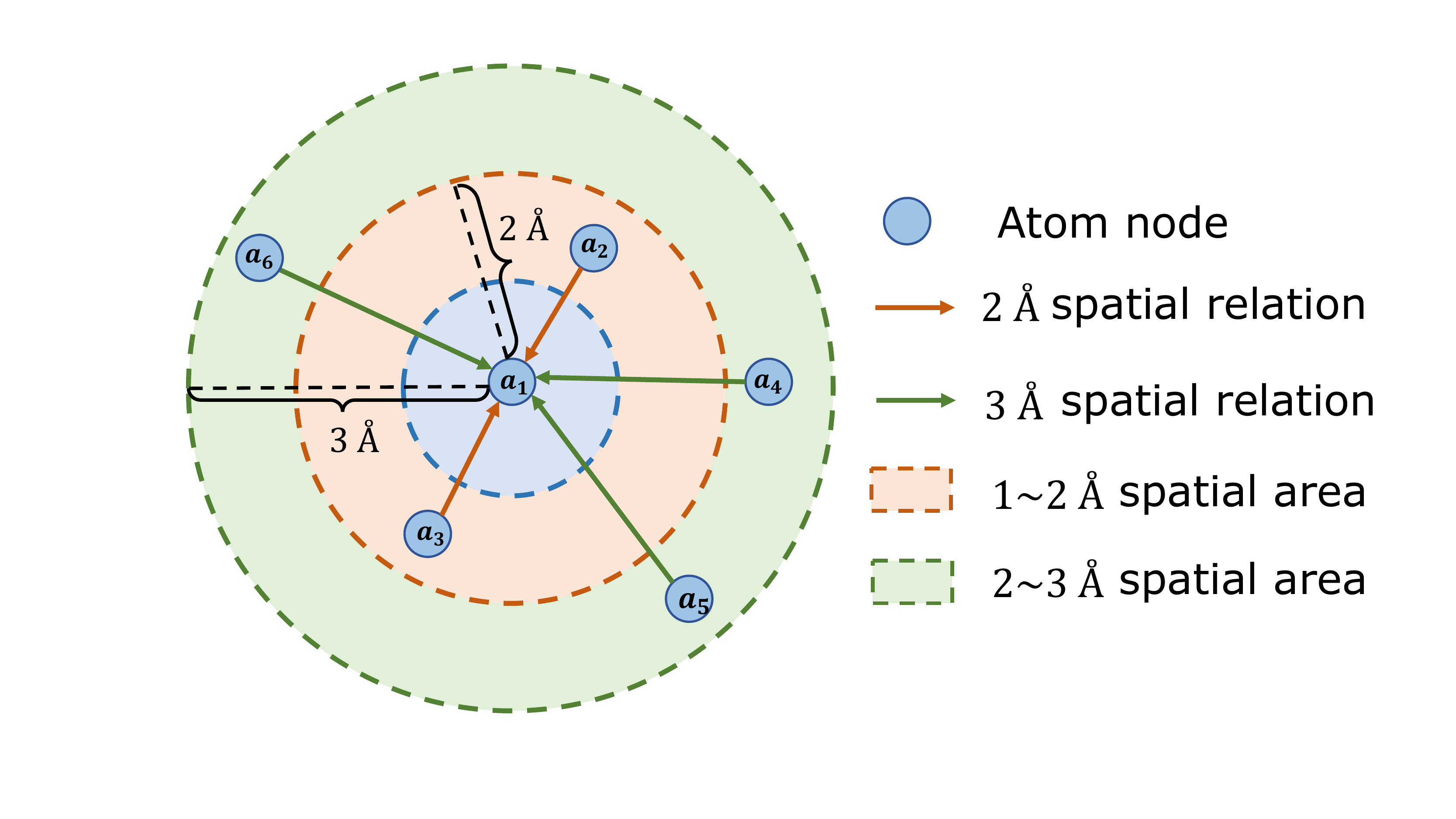}
\vspace{-5mm}
\caption{A toy example of spatial position encoding.}
\vspace{-3mm}
\label{pos-enc}
\end{figure}

To address the aforementioned limitations of CNN and GNN models, in this paper, we propose a novel end-to-end learning framework for DTA prediction, namely di\underline{S}tance-aware \underline{M}olecule Graph \underline{A}ttention \underline{N}etwork (\model). Figure \ref{model-frame} demonstrates the framework of our proposed model. Firstly, due to the lack of necessary spatial information in the basic molecular graph, our model takes the \underline{S}patial-enhanced \underline{Po}cket-ligand \underline{G}raph (SPoG) as input, which is constructed on the basis of protein pocket and ligand binding pose as shown in Figure \ref{lig-pock}(c). SpoG involves not only the 3D spatial distance information but also interactions between ligand and protein. The 
definition and construction of SPoG will be introduced in section \ref{sec-pre}. Then we design a spatial position encoding mechanism to model different spatial distance relations among atoms as illustrated in Figure \ref{pos-enc}.  Since the relative spatial distance information of atoms is attached to the edges upon the position encoding mechanism, it is difficult to propagate such information by aggregating nodes directly. Thus, we further invent a hierarchical attentive structure which has two steps: edge-level aggregation and node-level aggregation. This hierarchical structure can first capture spatial dependencies in edge-level stage, and then distinguish multiple spatial relations of 3D structure in node-level stage. To summarize, the main contributions of our work are as follows:

\begin{itemize}
    \item We first study the problem of DTA prediction by constructing a unique spatial-enhanced pocket-ligand graph and propose a novel distance-aware molecule graph attention network named {\model}.
    \item The \model employs both edge-level and node-level attentive aggregation with leveraging the spatial distance information and relative position of atoms. To the best of our knowledge, we are the first to adopt the hierarchical GNN structure with distance-ware attention for DTA prediction.
    \item The experiments on two datasets demonstrate that the proposed model outperforms the classic baselines and state-of-the-art GNN methods, {which shows great application potential for drug discovery.}
\end{itemize}

%\zhoucom{there is no any description about Figure \ref{sp-dist}?}

%% file: section/sec2-related.tex
\section{Related Work}
\label{sec-related}

Since our study %research work 
is dedicated to predicting the drug–target binding affinity with designing a new GNN architecture,
%graph neural network, 
%in this section, 
we briefly introduce the related work on these two topics.

% {\bfseries Prediction of drug–target binding affinity.}
\subsection{Prediction of drug–target binding affinity}
% 1. simulation methods
Drug-target binding affinity (DTA) prediction in drug discovery has attracted a large number of researchers' interest. Many previous works focused on simulation-based methods\cite{wang2003comparative} or classic machine learning models\cite{ballester2014does}, with requiring external expert domain knowledge. 
% 2. DL methods
Recently, most of the existing studies aim at solving the DTA problem based on deep learning models, such as DeepDTA\cite{ozturk2018deepdta} and WideDTA\cite{ozturk2019widedta}. These models utilize 1D convolutions and pooling to capture potential patterns from 1D ligand sequence and protein sequence. Thus, the necessary spatial and structure information \hidespace{is not taken into account}is neglected\hidespace{ in these models}. The recent advanced GNN models to incorporate the structure information of drug-target complex, like GraphDTA \cite{nguyen2020graphdta}, has shown better performance than them. Therefore, we did not compare with such 1D  convolution methods in our experiments.
% 3. 3D-CNN methods
There are also some works\cite{stepniewska2018development} constructing 3D image from drug-target complexes to use 3D convolutions (3D-CNN) to take advantage of spatially-local correlation. Though such a 3D-CNN
approach can learn spatial information, it has potential drawbacks. On the one hand, 3D-CNN requires a large number of model parameters, but the size of training data is limited for DTA problem.
On the other hand, the position of ligand or protein in different complexes is changeable, such as different angle rotation, which means the spatial structure of 3D image modeling is inevitably incomplete. To better learn the relative spatial information, our work develops the GNN architecture with integrating position of atoms for DTA prediction.

% 4. Other works -- binary classification mode, such as DrugQA and GNN_DTI(a distance-aware GNN) ...

% \B{Graph Neural Networks.}
\subsection{Graph Neural Networks}
With the increasing popularity of graph neural networks (GNNs), much attention has been devoted to applying GNN for molecular representation learning.
% 4. GNN methods
To integrate topological structure of molecular graph, GraphDTA\cite{nguyen2020graphdta} adopts several powerful GNNs\cite{kipf2017semi,velivckovic2018graph,xu2018powerful} to learn the drug presentation. Although GraphDTA shows reasonable performance in DTA prediction, it lacks the ability of learning position information in molecular graph and interaction information between drug-target. By contrast, our work offers a new perspective for drug-target prediction with the assistance of the critical relative position in molecular graph. There are %\li{also some} 
only a few of works studying GNNs with considering spatial information in recent years. MGCN\cite{lu2019molecular} and DimeNet\cite{klicpera2020directional} utilize the classic radial basis function to combine the meaningful distance information on the original graph. However, only the covalent bonding correlation is not sufficient for 3D
structure learning. What's more, different distances between atoms indicate different relations, while RBF can not provide this information explicitly. 
It is also worth noting that all these models fail to aggregate the spatial information attentively, and they are designed for molecular property prediction, which is quite different from \hidespace{the} DTA prediction\hidespace{problem}. 
To the best of our knowledge, we are the first to propose a dedicated GNN model tailored to the DTA problem which can identify multiple spatial relations while aggregating with distance-aware attention mechanism.

%Our proposed model can identify multiple spatial relation while aggregating with position-aware attention mechanism.

% a neglected deficiency is that they model DTA as an independent data sample and do not consider their multiple related correlations 

%% file: section/sec3-preliminaries.tex
\section{Preliminaries}
\label{sec-pre}
In this section, we first formally introduce the drug-target binding affinity (DTA) prediction problem. Then we describe the construction process of the spatial-enhanced pocket-ligand graph in detail. 

\subsection{Problem Statement}

We first clarify several related concepts of DTA prediction. The \textbf{drug} is referred to as a chemical compound, which is also called \textbf{ligand} in DTA prediction. What's more, the \textbf{protein} is also called \textbf{target}. Thus, in this paper, the main objects of drug-target binding affinity prediction are related to \textbf{ligand} and \textbf{protein}. As illustrated in Figure \ref{lig-pock}, protein-ligand interactions can be interpreted in two ways. We will introduce and compare them as follows.

Given a drug compound and a target protein, the DTA prediction task is to predict the binding affinity between them. In general, we use $L$ and $P$ to represent the input drug (or ligand) and the input target (or protein). Both can be a graph, a sequence, or other format input. The predicted binding affinity $y$ is a continuous real number value. Traditionally, DTA prediction can be defined as a regression task:
\begin{linenomath}
\begin{equation}
\label{df-eq-1}
    f:(L, P) \rightarrow y
\end{equation}
\end{linenomath}

To overcome the limitation of weak interaction information between drug and protein when splitting them into two-part input, as illustrated in Figure \ref{lig-pock}(c), a more appropriate formulation is to represent drug-protein complex as protein-ligand binding pose \cite{lim2019predicting} with preserving the essential spatial structure, we call it \textit{pocket-ligand} for short in this paper. Meanwhile, the size of pocket-ligand is significantly less than that of the original ligand-protein. The pocket-ligand graph can be denoted as $\mathcal{G}$, and the distance matrix of atoms in graph $\mathcal{G}$ can be represented as $D$. The construction of $\mathcal{G}$ and $D$ will be introduced next.
%and the construction of $\mathcal{G}$ will be introduced next. Here we use $S$ to represent the position matrix of atoms in graph $\mathcal{G}$. 
Now the problem can be defined as:
\begin{linenomath}
\begin{equation}
\label{df-eq-2}
    g:(\mathcal{G}, D) \rightarrow y
\end{equation}
\end{linenomath}

% \begin{figure}
%   \centering
%   \subfigure[Distance]{
%     \label{component-exp-b} %% label for second subfigure
%     \includegraphics[width=0.48\columnwidth]{figures/distance_distribution.pdf}}
%       \subfigure[Size]{
%     \label{component-exp-a} %% label for first subfigure
%     \includegraphics[width=0.48\columnwidth]{figures/ligand_pocket_affinity_b.pdf}}
%   \vspace{-3mm}
%   \caption{Examples.}
%   \label{component-exp} %% label for entire figure
% \end{figure}

% \begin{figure}
% \centering
% \includegraphics[width=1.0\columnwidth]{figures/distance_distribution.pdf}
% \vspace{-4mm}
% \caption{Size and binding affinity.}
% \label{Size}

%\begin{figure}
%  \centering
%  \subfigure[]{
%    \label{sp-max-dist} %% label for second subfigure
%    \includegraphics[width=0.46\columnwidth]{figures/ligand_pocket_affinity_b-x.pdf}}
%      \subfigure[]{
%    \label{sp-distribution} %% label for first subfigure
%    \includegraphics[width=0.49\columnwidth]{figures/distance_distribution-1.pdf}}
%  \vspace{-3mm}
%  \caption{Spatial distance visual analysis.}
%  \vspace{-3mm}
%  \label{sp-dist} %% label for entire figure
%\end{figure}
%%\zhoucom{there is no any description about Figure \ref{sp-dist}?}

%\begin{figure}
%\centering
%\includegraphics[width=1.0\columnwidth]{figures/pos-enc-1.pdf}
%\vspace{-8mm}
%\caption{A toy example of spatial position encoding.}
%\vspace{-3mm}
%\label{pos-enc}
%\end{figure}

% \begin{figure}
% \centering
% \includegraphics[width=1.0\columnwidth]{figures/pos-enc-1.pdf}
% \vspace{-8mm}
% \caption{A toy example of spatial position encoding.}
% \vspace{-3mm}
% \label{pos-enc}
% \end{figure}

\subsection{Pocket-ligand Graph Construction}
As we claimed in section \ref{sec-related}, the \hide{spatial}structure information in the original molecular graph with only the covalent bonding correlation is not enough. More spatial edges should be included in the graph to provide adequate 3D structure information. What's more, there is no natural bonding between ligand and protein. Therefore, the input interaction graph of our proposed model is the spatial-enhanced pocket-ligand graph (\graph). We denote the \graph  by a new graph $\mathcal{G}=(V=V_M \cup V_P, E)$, where $V=\{a_1, a_2, ..., a_N\}$, $V_M$ and $V_P$ are the atom set of ligand and protein pocket. To build spatial-enhanced edges for $\mathcal{G}$, we first calculate the spatial distances between all atoms in 3D space, the distance matrix is denoted as $D$.  Then a threshold $\theta_{d}$ is set to preserve the correlation edge $e_{ij}$ between a pair of atoms. In this way, the edge set can be built:
\begin{linenomath}
\begin{equation}
    E = \{e_{ij}=(v_i,v_j) | v_i,v_j \in V, D_{ij} \leq \theta_d \}
\end{equation}
\end{linenomath}

%% file: section/sec4-method.tex
\section{Our Proposed Model}
\label{sec-method}
In this section, we propose a distance-aware molecule graph attention network (\model) to address the drug-target binding affinity prediction problem.

\begin{figure}[t]
\centering
\includegraphics[width=0.5\textwidth]{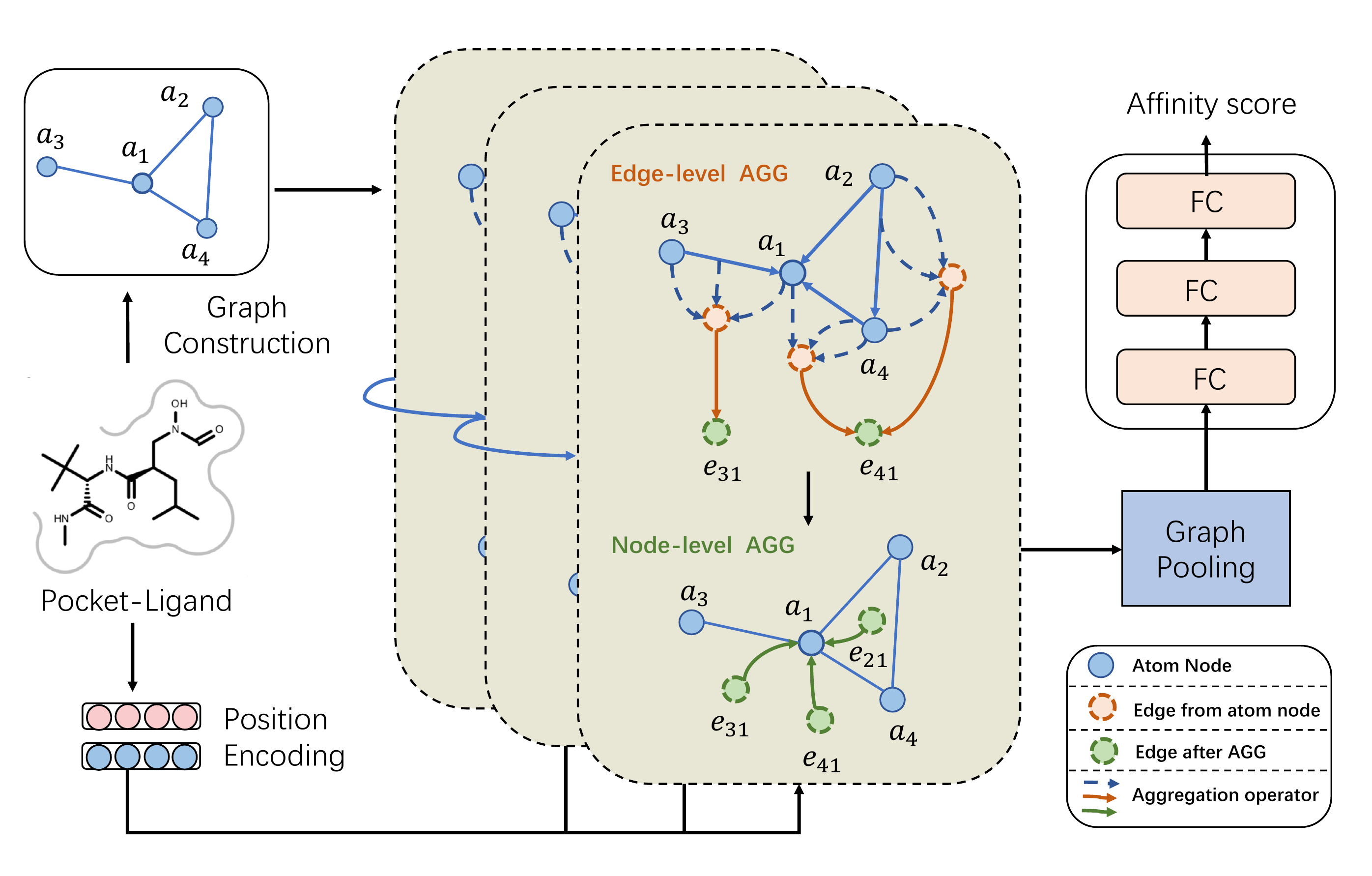} %
%\vspace{-5mm}
\caption{Illustration of Distance-aware Molecule Graph Attention Network (\model) for DTA prediction.
}
\label{model-frame}
\vspace{-1mm}
\end{figure}

\subsection{Overview}
Figure \ref{model-frame} depicts the framework of \model. As mentioned in Section \ref{sec-pre}, it takes the spatial-enhanced pocket-ligand graph $\mathcal{G}$ and position matrix $S$ as input. \model contains distance-aware molecule graph attention layers designed for DTA prediction, which propagates the atom representation spatially and attentively. The two aggregation operations, edge-level aggregation and node-level aggregation, play a synergistic effect on improving the performance. Then the graph pooling layer produces the graph representation to obtain the final binding affinity score by applying several fully connected layers. In the following sections, we use $a_i$ and $\bm{a}_{i}$ (i.e. bold letter) to represent the atom node and the embedding of atom $i$ respectively. Similarly, the edge ($a_i$,$a_j$) is denoted as $e_{ij}$ and the embedding of edge is denoted as $\bm{e}_{ij}$.

\subsection{Distance-aware Molecule Graph Attention}\label{sec:swmga}
Different from the general graph, the molecular graph, like drug compound and protein, has a unique 3D spatial structure, which may affect the molecular property and interaction strongly. For example, as shown in Figure \ref{sp-max-dist} and \ref{sp-distribution}, the binding affinity is related closely to the max distance in complexes. Also, the spatial distance distribution in pocket-ligand indicates that the covalent bond can be formed when the spatial distance between atoms is less than a certain distance. The ability to learn such spatial structure and position information is critical to biological modeling, especially in the DTA prediction problem. To integrate the topological structure and spatial position information suitably, a novel distance-aware molecule graph attention network (\model) is designed with a hierarchical attentive aggregation structure: edge-level aggregation $\rightarrow$ node-level aggregation. First, we adopt an edge-level aggregation to deliver the pairwise atoms' embedding with position information to get the edge embedding. After that, node-level aggregation can get the optimal weighted combination of the edge embeddings for each atom node with a distance-aware attention.

% \begin{figure}
% \centering
% \includegraphics[width=1.0\columnwidth]{figures/pos-enc-1.pdf}
% \vspace{-8mm}
% \caption{A toy example of spatial position encoding.}
% \vspace{-3mm}
% \label{pos-enc}
% \end{figure}

\subsubsection{Position Encoding.}
The position of atoms in pocket-ligand is defined by 3D coordinates, forming the input position matrix $S \in \mathbb{R}^{N \times 3}$. Considering the variability of coordinates that are manually defined, we convert $S$ into a relative spatial matrix, that is distance matrix $D \in \mathbb{R}^{N \times N}$. From Figure \ref{sp-distribution}, we noticed that spatial distance indicates different meaningful correlations between atoms. 
Therefore, the spatial information is encoded by applying a one-hot encoder to split the scalar distances into $b$ buckets, leading to a multiple position relation matrix $D^R \in \mathbb{R}^{N \times N \times b}$. Taking \ref{pos-enc} as an example, we divide neighbors of $a_1$ into different spatial relations. Then we adopt a dense layer to obtain the position embedding $\bm{p}_{ij}$ for each pocket-ligand edge $e_{ij}$.
\begin{linenomath}
\begin{equation}
   \bm{p}_{ij} = W_p D^R_{ij}
\end{equation}
\end{linenomath}

Where $W_p \in \mathbb{R}^{d \times b}$ is the transformation weight matrix. Next, we take the position embedding into the hierarchical aggregation layer for both edge-level and node-level.

\subsubsection{Edge-level Aggregation.}
Considering the relative position information is attached to a pair of atoms, the key challenge of applying GNN to learning position information is how to propagate such pairwise information on a molecular graph. To this end, \gnn first employs an attentive aggregation for edges to capture long-range dependencies among atom nodes. In this sense, we introduce the definition of the edge neighbors, denoted as $\mathcal{N}_e$. If there is a path: $a_k \rightarrow a_i \rightarrow a_j$, the edge $e_{ki}=(a_k,a_i)$ is a neighbor of the edge $e_{ij}=(a_i,a_j)$, denoted as $e_{ki} \rightarrow e_{ij}$, more formally: 
\begin{linenomath}
\begin{equation}
    \mathcal{N}_e(e_{ij}) = \{e_{ki} |e_{ki} \in E, k \ne j \}
\end{equation}
\end{linenomath}

As illustrated in Figure \ref{model-frame}, the edge embedding $\bm{e}^l_{ij}$ is firstly updated through the node-to-edge aggregation:
\begin{linenomath}
\begin{align}
\label{n2e-agg}
    \bm{e}_{ij}^l & = AGG_{node \rightarrow edge} (a_i, a_j, p_{ij})\nonumber \\
   &= \sigma(W_{ne}^l \cdot [\bm{a}_i^{l-1} \oplus \bm{a}_j^{l-1} \hide{\operatornamewithlimits{||}} \oplus \bm{p}_{ij}]+b^l_{ne}) 
\end{align}
\end{linenomath}

Where $\oplus$ is the concatenation operation over two vectors, $W_{ne}^l$ is the transformation matrix at the $l$-th layer to combine node embedding and position embedding, $b^l_{ne}$ is the bias vector, and $\sigma$ is the activation function for non-linearity. For each edge $e_{ij}$, inspired by the GAT \cite{velivckovic2018graph}, the following attentive aggregation over edge neighbors is formulated as:
\begin{linenomath}
\begin{align}
    \bm{e}^{l}_{ij} & = AGG_{edge \rightarrow edge} (e_{ij}, \mathcal{N}_e(e_{ij}))\nonumber \\ &=\sum_{e_{ki} \in \mathcal{N}_e(e_{ij})} \alpha_{k,i,j}^l W_e^l \bm{e}_{ki}^l 
\end{align}
\end{linenomath}

where $W_e^l$ is the weight matrix and $\alpha_{k,i,j}^l$ is the normalized attention weight of edge neighbor $e_{ki}$ via the softmax function:
\begin{linenomath}
\begin{equation}
    \alpha_{k,i,j}^l = \frac{exp(\sigma_a(\bm{a}_{e,l}^T[W_e^l\bm{e}_{ij}^l \oplus W_e^l \bm{e}_{ki}^l]))}{\sum_{e_{ti} \in \mathcal{N}_e(e_{ij})} exp(\sigma_a(\bm{a}_{e,l}^T[W_e^l\bm{e}_{ij}^l \oplus W_e^l \bm{e}_{ti}^l]))}
\end{equation}
\end{linenomath}

where $\bm{a}_{e,l}$ is the attention parameter for measuring weights of edge neighbors and we use LeakyReLU as the activation function. Thanks to the position information injection in equation \ref{n2e-agg} after node-to-edge aggregation, the later attentive edge-to-edge aggregation can acquire the long-range dependencies adaptively in molecular graph.

\subsubsection{Node-level Aggregation.}
Furthermore, after obtaining the edge embedding $\bm{e}^l_{ij}$ from edge-level aggregation, we apply the node-level aggregation to fuse the position-enhanced information with the capability of discriminating multiple spatial relations among atoms. By means of aggregating all related edges for each node with the specially designed distance-aware attention, the combination of edge representations involves the necessary spatial and topological structure information. Similar to edge neighbors set $\mathcal{N}_e$, we define the edge neighbors of a node $a_i$ as follows:
\begin{linenomath}
\begin{equation}
    \mathcal{N}_{eon}(a_i) = \{e_{ki} |e_{ki}=(a_k,a_i) \in E \}
\end{equation}
\end{linenomath}

We first convert the edge embedding and atom node embedding into the hidden representation $\bm{h}_{k,i,e}$ and $\bm{h}_{i,a}$ in the same vector space by performing a linear transformation:
\begin{linenomath}
\begin{equation}
    \bm{h}^l_{k,i,e} = W_h^l \bm{e}_{ki}^l, \ \ \bm{h}^l_{i,a} = W_h^l \bm{a}_{i}^{l-1}
\end{equation}
\end{linenomath}

Where $W^l_h$ is the weight matrix in $l$-th layer. Then we propose the distance-aware attention to learn the weight among multi-relation edges.
The importance of the edge $e_{ki}$ for destination atom $a_i$ can be formulated as follows:
\begin{linenomath}
\begin{align}
    w_{ki}^l & = attn_{eon} (a_i, e_{ki}, p_{ki})\nonumber \\
   &= \sigma(\bm{a}_{n,l}^T\cdot[\bm{h}^l_{i,a} \oplus \bm{h}^l_{k,i,e} \oplus W^l_s\bm{p}_{ki} ])
\end{align}
\end{linenomath}

where $\bm{a}_{n,l}$ is the node attention parameter for measuring weights of edge neighbors, $W^l_s$ is the weight matrix for position transformation and $\sigma$ is the activation function. Then the softmax function is used for normalization: 
\begin{linenomath}
\begin{equation}
    \beta_{ki}^l = \frac{exp(w_{ki}^l)}{\sum_{e_{ki} \in \mathcal{N}_{eon}(a_i)} exp(w_{ki}^l) }
\end{equation}
\end{linenomath}

The updated atom node embedding is calculated by the edge-to-node aggregation. We also develop the position-aware attention to multi-head attention version as GAT for better stability with $M$ independent attention mechanisms: 
\begin{linenomath}
\begin{align}
    \bm{a}^{l}_{i} & = AGG_{edge \rightarrow node} (a_i, \mathcal{N}_{eon}(a_i))\nonumber \\ 
    &= \sigma \Big( \frac{1}{M} \sum_{m=1}^M\sum_{e_{ki} \in \mathcal{N}_{eon}(a_i)} \beta_{ki}^{l,m} \bm{h}_{k,i,e}^{l,m} \Big)
\end{align}
\end{linenomath}

As shown in Figure \ref{model-frame}, we further stack $L$ position-aware Molecule GAT layers to learn more adequate position and structure information for drug-target binding affinity prediction, and we use $\bm{a}_i=\bm{a}^L_i$ to represent the final embedding of atom $a_i$.

\subsection{Drug-target Binding Affinity Prediction}
After performing \gnn layers, we obtain the representations of atoms in pocket-ligand. We can employ a graph pooling layer for all atoms to get the global pocket-ligand embedding. In our study, the graph-level representation $\bm{g}$ is calculated by summation:
\begin{linenomath}
\begin{equation}
    \bm{g} = \sum_{a_i \in V} \bm{a}_i
\end{equation}
\end{linenomath}

Then we feed $\bm{g}$ into fully connected layers to predict the drug-target binding affinity score:
\begin{linenomath}
\begin{equation}
    \hat{y} = W_o(\sigma(W_2 \sigma(W_1 \bm{g} + b_1))+b_2)+b_o)
\end{equation}
\end{linenomath}

We use the Mean Square Error (MSE) between the predicted valur $\hat{y}$ and the observed binding affinity $y$ as the loss function to train the model \model over all pocket-ligand complexes in dataset $\mathcal{D}$:
\begin{linenomath}
\begin{equation}
    \mathcal{L} = \frac{1}{|\mathcal{D}|}\sum_{(M,P) \in \mathcal{D}}(y-\hat{y})^2
\end{equation}
\end{linenomath}

%% file: section/sec5-experiment.tex
\section{Experiments}
In this section, we first introduce the datasets and experiment settings. Then we compare our proposed \model with other methods to predict drug-target binding affinity on two PDBbind datasets.

\subsection{Datasets}
As our proposed model takes advantage of 3D positions of atoms, the experimental dataset is required to provide such spatial information. So we conducted experiments using two public released PDBbind datasets (v.2016 and v.2019) to evaluate the effectiveness of \model and baselines.

\subsubsection{PDBbind.} The PDBbind\footnote{http://www.pdbbind-cn.org/download.php} dataset \cite{wang2005pdbbind} is a well-known dataset for predicting the binding affinity of drug-target complexes, which is composed of 3D structures of molecular complexes and the corresponding experimentally determined binding affinities expressed with $pK_a$ values. Each dataset has three subsets: \textit{general set}, \textit{refined set} and \textit{core set}. The \textit{general set} contains all complexes with relatively lower quality, while the \textit{refined set} is a subset of the \textit{general set} with higher quality, which is used as the training set in our experiment. The \textit{core set} is designed as the highest quality benchmark, and it is usually used as a test set. With regard to the most used PBDBind v.2016, there are 4057 complexes in the \textit{refined set} and 290 complexes in the \textit{core set}. Besides, we also use the latest released PDBbind v.2019 with 4852 and 285 complexes in these two subsets, which is updated on the previous v.2016 edition. 
% \subsubsection{Features.} 

% \begin{table}[t]
% 	\caption{Statistics of two datasets.}
% 	%\vspace{-3mm}
% 	\label{table-dataset}
% 	\centering
% 	\begin{tabular}{cccccc}
% 		\toprule
% 		Dataset	&	\#training set	&	\#core set	&	\#Pairs	&   \#Heat Maps    \\
% 		\midrule
% 		PDBBind v.2016	&	96,972  &	1,113,962	&	18,731	&	19,841	\\
% 		PDBbind v.2019	&	32,449	&	256,954	&   7,514	&	9,624	\\
% 		\bottomrule
% 	\end{tabular}
% \end{table}

\subsection{Evaluation metrics}
To comprehensively evaluate the model performance, we use Root Mean Square Error (RMSE) and Mean Absolute Error (MAE) to measure the prediction error. The performance of a model is also quantitatively evaluated by the classic Pearson's correlation coefficient (R) and the standard deviation (SD) in regression to measure the linear correlation between predictions and the experimental \hidespace{binding} constants. As introduced in CASF \cite{li2014comparative}, SD is defined as follows:
\begin{linenomath}
\begin{equation}
    SD = \sqrt{\frac{1}{|\mathcal{D}|-1}\sum_{i=1}^{|\mathcal{D}|} [y_i - (a+b \hat{y}_i)]^2}
\end{equation}
\end{linenomath}

where $\hat{y}_i$ and $y_i$ respectively represent the predicted and experimental value of the $i$-th complex in dataset $\mathcal{D}$, and $a$ and $b$ are the intercept and the slope of the regression line, respectively.

\subsection{Methods for Comparison}
We compare our \model model with the following methods to predict the drug-target binding affinity:
\begin{itemize}
\item \B{LR} uses linear regression for drug-target binding affinity prediction. We calculate the inter-molecular interaction features introduced in \cite{ballester2010machine} as the input and predict the affinity scores.
\item \B{SVR} is a variant of support machine vector (SVM) for regression task. We use the same features as LR. Please note that these strong graph-level features are extracted by domain knowledge with considering the interaction and spatial information among atoms, which are time-consuming. %time-costing.
\item \B{Pafnucy} \cite{stepniewska2018development} is a 3D CNN model designed to learn the spatial structure of protein-ligand complexes for drug-target binding affinity prediction.

%designed for drug-target binding affinity prediction, which learns the spatial structure of protein-ligand complexes.
\item \B{GraphDTA} \cite{nguyen2020graphdta} is an effective graph neural network model, which introduced GNN into DTA prediction. The graph with atoms as nodes and bonds as edges is constructed to describe drug molecules. It also uses CNN to learn the protein sequence representation. There are four variants with different GNN models:  \textit{Graph-GCN}, \textit{Graph-GIN}, \textit{Graph-GAT} and \textit{Graph-GCN+GAT}.
% \B{Graph-GCN}, \B{Graph-GIN}, \B{Graph-GAT} and \B{Graph-GCN+GAT}.
\item \B{\graph-DTA} improves the GraphDTA by inputting our spatial-enhanced pocket-ligand graph (\graph) instead of the drug molecular graph into the GNN model. We name the four variants as \textit{\graph-GCN}, \textit{\graph-GIN}, \textit{\graph-GAT} and \textit{\graph-GCN+GAT}.
\item \B{\model-NoEdge} only performs the node-level aggregation on the \graph. The edge-level aggregation stage is removed and the node embedding is updated from the node neighbors of each atom. 
\item \B{\model-NoSpAttn} replaces the distance-aware attention in our model by general graph attention without spatial information while conducting node-level aggregation.
\end{itemize}

\subsection{Implementation Details}
\subsubsection{Settings.} We randomly pick 90\% from each \textit{refined set} in PDBbind v.2016 and v.2019 as the training datasets, and the remaining 10\% complexes are used for validation. The main statistics of two PDBbind datasets are summarized in Table
\ref{table-dataset}. %Our model is implemented with PaddlePaddle\footnote{https://github.com/PaddlePaddle/Paddle}. 
We optimize models with Adam optimizer, where the batch size is fixed at 32. Besides, we construct the spatial-enhanced pocket-ligand graph with the threshold $\theta_d=5 \ \si{\angstrom}$ to ensure the appropriate size of graph. For \model, we set the dimension of atom features as 36, and we keep the same dimension for node embedding and edge embedding. The hidden size is set to 128. For position encoding, we divide the spatial distances into 4 buckets. We set the learning rate to $5e^{-4}$, the number of attention head $M$ to 4 and the dropout ratio to 0.2. For Pafnucy and GraphDTA models, we input the same 36-dimension atom features as \model. For all baseline models, we use default optimal parameter settings as in their original implementations.

\begin{table}
	\centering
	\begin{tabular}{cccccc}
		\toprule
		Dataset	&	Training	&	Validation	&	Testing (\textit{core set}) \\
		\midrule
		v.2016	&	3,390  & 377	& 290 	\\
		v.2019	&	4,127	& 459	&  285	\\
		\bottomrule
	\end{tabular}
		\caption{Statistical complexes in two PDBbind datasets.}
	\vspace{-3mm}
	\label{table-dataset}
\end{table}

\subsubsection{Features.} For 3D-CNN and GNN models, the atom features used according to \cite{stepniewska2018development} include atom type and hybridization, the numbers of bonds with other heavy-atoms and hetero-atoms, atom properties such as hydrophobic, and partial charge. In total, 18 features are used to describe an atom. Considering the heterogeneity in the pocket-ligand graph, we further extend atom features to a 36-dimension vector, where the 1st to 18th elements represent ligand atoms and the 19th to 36th elements represent protein atoms.

\input{section/tbl-experiment.tex}

\subsection{Performance Evaluation on PDBbind dataset}

% \begin{figure}
% \centering
% \includegraphics[width=1.0\columnwidth]{figures/model_prediction.pdf}
% \vspace{-4mm}
% \caption{Predictions for PDBbind v.2016}
% \label{prediction-2016}
% \end{figure}

We compare our model with the baseline models mentioned above in two PDBbind datasets v.2016 and v.2019. The experimental results reported in Table \ref{table-expriemnt} are obtained over five runs repeatedly, and the mean value is calculated as well as the standard deviation in parentheses. We first evaluate our model with the previous works, and then analyze the effectiveness of the injected spatial information in our model.

\subsubsection{Predictive performance.}
As shown in Table \ref{table-expriemnt}, \model significantly outperforms the baselines in all metrics across the two datasets. More specifically, Pafnucy achieves relatively poor results, which indicates the limitation of the 3D-CNN model. As we have mentioned in Section \ref{sec-related}, although 3D-CNN models can learn the spatial information by treating the protein-ligand complexes as images, it's likely that the positions of atoms are influenced by the rotation and translation of the coordinate system. It might make the model confused when learning binding structures in different complexes. As a result, the 3D-CNN model can only capture restricted position information. Benefit from the strong features of the occurrence for atom types within the specified spatial distance, the performance of SVR is better than LR and Pafnucy. Due to the ability of aggregating information of spatial position and topological structure, our model has much better performance than the above baselines.

For graph neural networks, the GNN models of \graph-DTA perform better than GraphDTA on the whole, demonstrating that the richer spatial information is helpful for DTA prediction. What's more, we observe that the GAT model achieves the most prominent performance improvement. The potential reason is that the attention mechanism helps to find out meaningful neighbors among added spatial relations. However, these GNN models fail to capture the spatial information, and our \model offers the average relative performance gain of 10.9\% in RMSE over the best baselines on DPBbind v.2016 dataset.

\subsubsection{Influence of spatial information.}
\begin{figure}
  \centering
  \subfigure[Results on PDBbind v.2016]{
    \label{exp-ablation-2016} %% label for second subfigure
    \includegraphics[width=0.35\columnwidth]{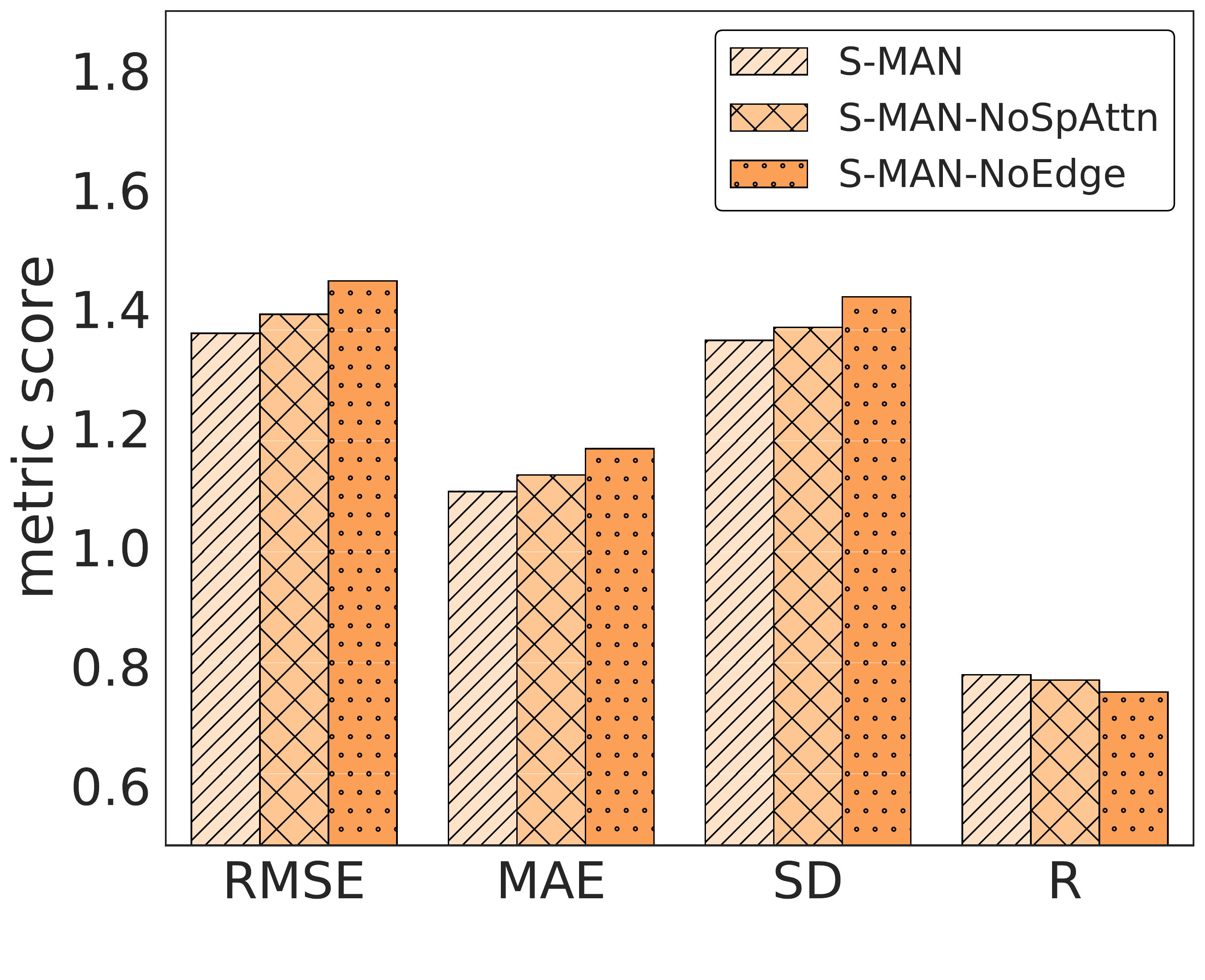}}
      \subfigure[Results on PDBbind v.2019]{
    \label{exp-ablation-2019} %% label for first subfigure
    \includegraphics[width=0.35\columnwidth]{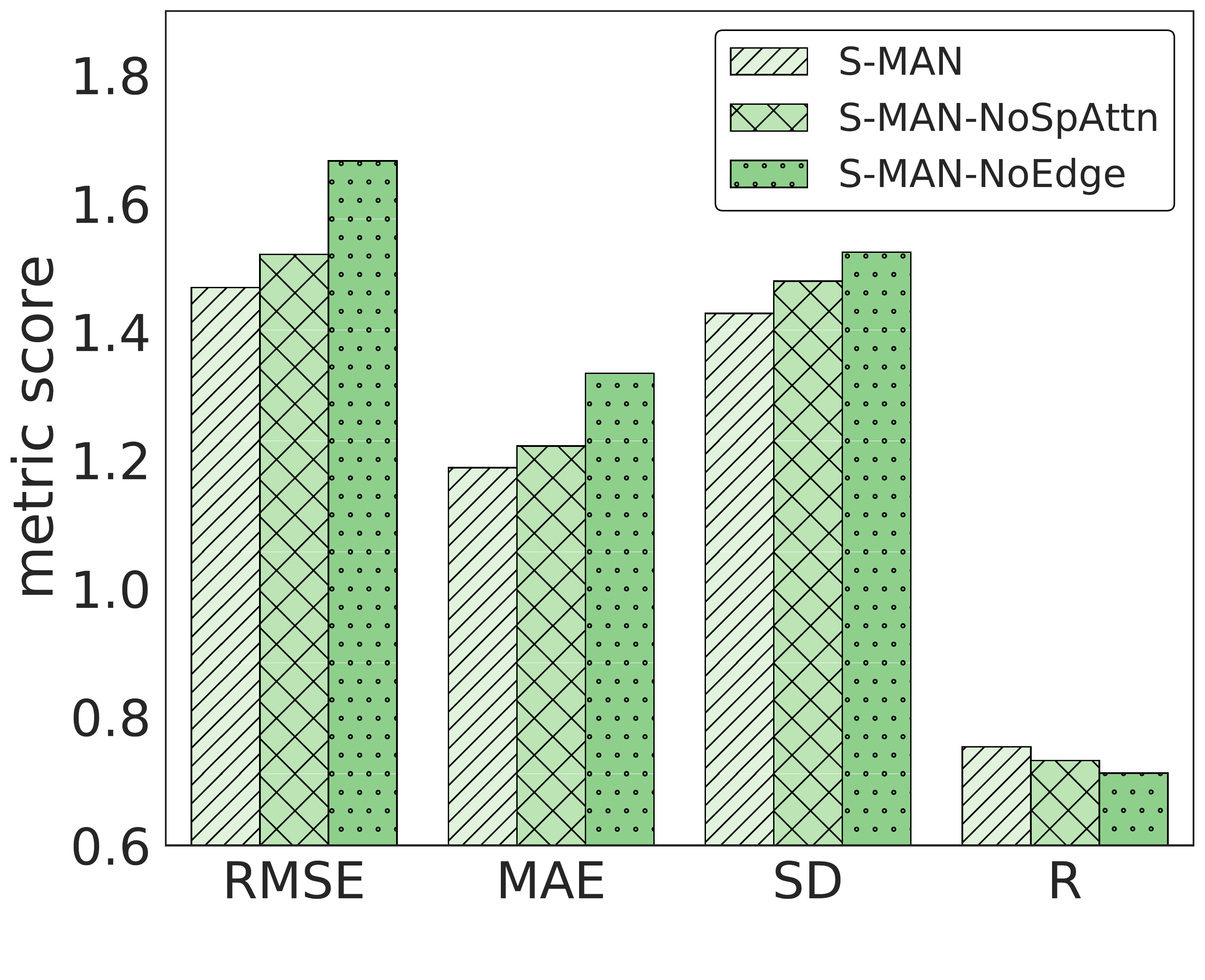}}
  \vspace{-2mm}
  \caption{Evaluation of \model with its variants.}
  \vspace{-2mm}
  \label{exp-ablation} %% label for entire figure
\end{figure}

To study the effectiveness of distance-aware attention and edge-level aggregation, we further conduct experiments for the variants of \model. As illustrated in Figure \ref{exp-ablation}, the results show that removing the edge-level aggregation degrades the model's performance, proving that the spatial information carried by the edge is critical to drug-target binding affinity prediction. Moreover, the prediction error of \model-NoEdge is higher than \model-NoSpAttn, it indicates that the edge-level aggregation plays a more significant role in our model, which demonstrates the necessity of the hierarchical structure. As \model-NoSpAttn ignores the position information of atoms and lacks the ability of identifying multiple spatial relations while executing the node-level aggregation, it performs worse than \model on both datasets.

\begin{figure}
  \centering
  \subfigure[]{
    \label{exp-bucket} %% label for second subfigure
    \includegraphics[width=0.35\columnwidth]{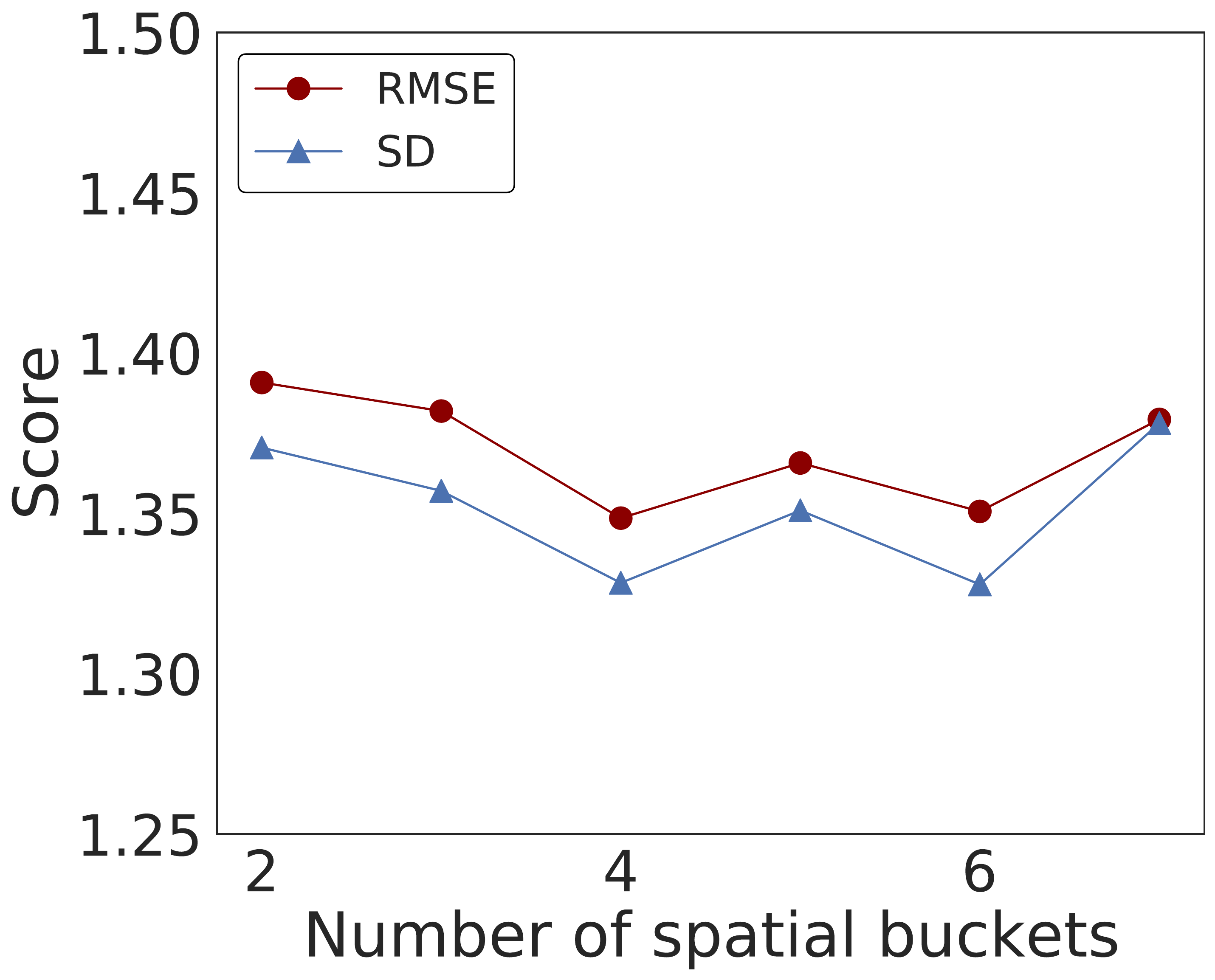}}
      \subfigure[]{
    \label{exp-layer} %% label for first subfigure
    \includegraphics[width=0.35\columnwidth]{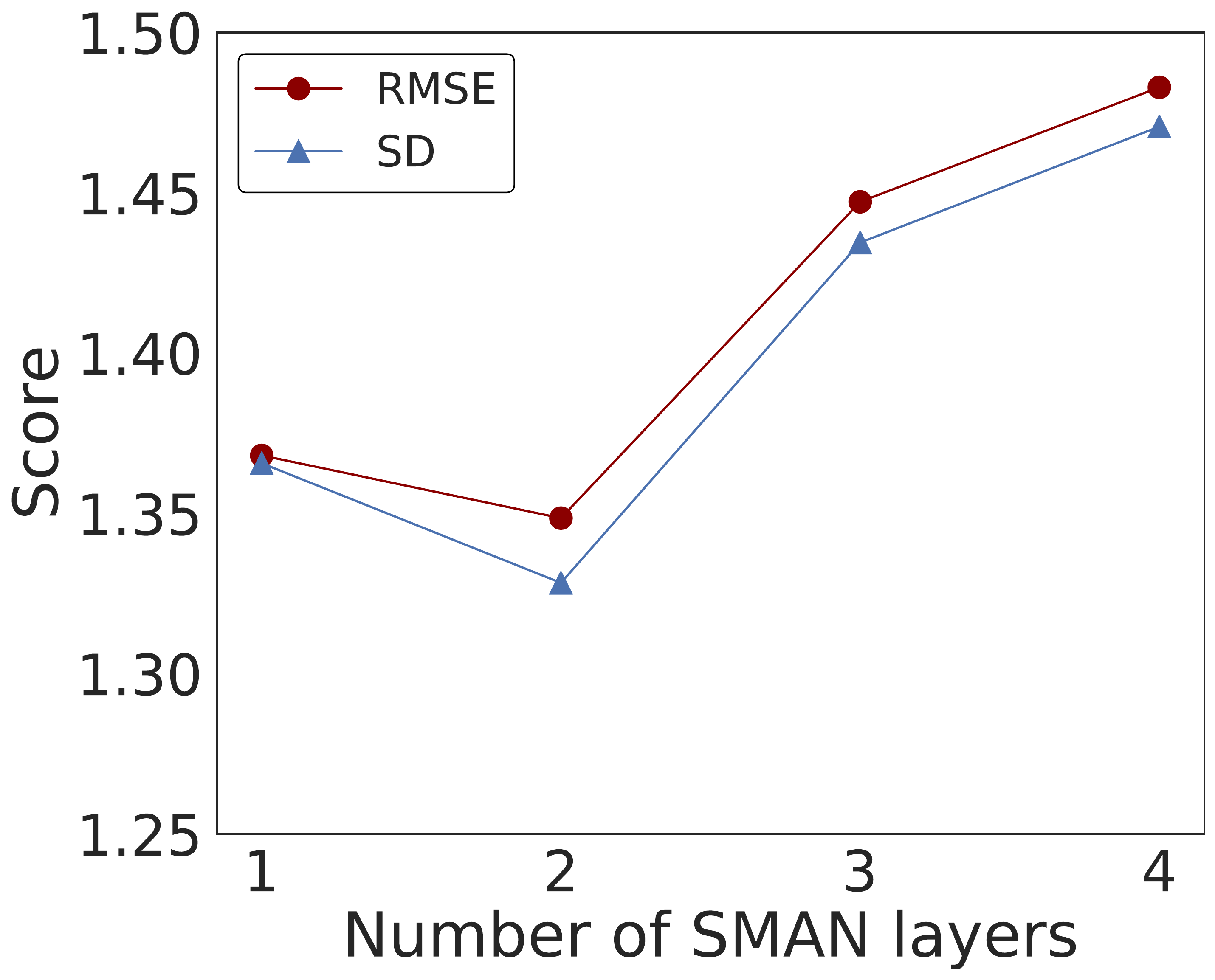}}
  \vspace{-2mm}
  \caption{Parameter sensitivity experiment results.}
  \vspace{-3mm}
  \label{exp-para} %% label for entire figure
\end{figure}

\subsection{Parameter Sensitivity Analysis} 

\subsubsection{Number of spatial buckets.}
To explore the impact of the spatial bucket setting parameter $b$, we conduct the parameter sensitivity experiment on the PDBbind v.2016 dataset by changing the number of spatial buckets. As shown in Figure \ref{exp-bucket}, when the parameter $b$ increases from 2 to 4, there are noticeable improvements on both two metrics, and the performance does not get better since $b > 4$. This is probably because more spatial relation information and position information is available for the model, while too many spatial buckets might produce unexpected noises.

\subsubsection{Number of \model layers.} 
We further study the influence of different numbers of \model layers by varying from 1 to 4. As shown in Figure \ref{exp-layer}, we observe that the performance of our model gets worse starting from 3 layers. This is because too many layers cause the overfitting of our model on the training set. It indicates that \model can achieve great performance with only one or two layers.

%% file: section/tbl-experiment.tex
\renewcommand\arraystretch{1.0}

\begin{table*}[t]
    \centering
	\scalebox{0.85}{\begin{tabular}{c|c|c|c|c|c|c|c|c}
		\toprule
		\multirow{2}{*}{} & \multicolumn{4}{c|}{PDBbind v.2016} & \multicolumn{4}{c}{PDBbind v.2019}   \\
		\cline{2-9}
		\rule{0pt}{10pt}
			&	RMSE	&	MAE	&	SD	& R &	RMSE	&	MAE	&	SD	& R	\\
		\midrule

		LR  & 1.677 (0.00) & 1.355 (0.00) & 1.605 (0.00) & 0.676 (0.00) & 1.693 (0.00)  &	1.374 (0.00)  & 1.620 (0.00)  & 0.667 (0.00)  \\
		SVR  & 1.562 (0.00) & 1.269 (0.00) & 1.496 (0.00) & 0.726 (0.00) & 1.577 (0.00) & 1.282 (0.00)  &	1.511 (0.00)  & 0.719 (0.00) \\
% 		XGboost & 1.483 (0.000) & 1.192 (0.000) & 1.437 (0.000) & 0.751 (0.000) & 1.487 (0.000) & 1.173 (0.000)  &	1.455 (0.000)  & 0.743 (0.000) \\
%         RF-Score    & 1.468 (0.005) & 1.186 (0.005)  & 1.352 (0.008)  & 0.783 (0.003) & 1.489 (0.003) & 1.193 (0.004) & 1.397 (0.004) & 0.766 (0.002) \\
		Pafnucy  & 1.601 (0.02) & 1.295 (0.02) & 1.584 (0.02) & 0.686 (0.01) & 1.907 (0.08)  &	1.520 (0.07)  & 1.711 (0.03)  & 0.617 (0.02) \\
		\midrule
        Graph-GIN  & 1.655 (0.04) & 1.248 (0.04) & 1.646 (0.04) & 0.654 (0.02) & 1.632 (0.03) & 1.238 (0.04) & 1.623 (0.03) & 0.665 (0.01)\\
		Graph-GCN  & 1.661 (0.04) & 1.278 (0.03) & 1.653 (0.03) & 0.650 (0.02) & 1.715 (0.04) & 1.304 (0.02) & 1.698 (0.03) & 0.624 (0.02) \\
		Graph-GAT  & 1.776 (0.04) & 1.378 (0.03) & 1.751 (0.03) & 0.593 (0.02) & 1.814 (0.01) & 1.396 (0.01) & 1.786 (0.01) & 0.570 (0.01) \\
		Graph-GCN+GAT & 1.539 (0.02) & 1.204 (0.02) & 1.537 (0.02) & 0.708 (0.01) & 1.646 (0.04) & 1.292 (0.04) & 1.642 (0.04) & 0.655 (0.02) \\
		\midrule
        \graph-GIN  & 1.663 (0.02) & 1.266 (0.01) & 1.646 (0.02) & 0.655 (0.01) & 1.713 (0.05) & 1.299 (0.04) & 1.693 (0.04) & 0.627 (0.02)	\\
		\graph-GCN  & 1.702 (0.04) & 1.292 (0.03) & 1.679 (0.04) & 0.636 (0.02) & 1.678 (0.02) & 1.288 (0.02) & 1.670 (0.02) & 0.640 (0.01) \\
		\graph-GAT  & 1.711 (0.02) & 1.310 (0.01) & 1.694 (0.02) & 0.628 (0.01) & 1.709 (0.01) & 1.304 (0.01) & 1.684 (0.00) & 0.632 (0.00) \\
		\graph-GCN+GAT & 1.526 (0.02) & 1.192 (0.02) & 1.526 (0.03) & 0.713 (0.01) & 1.548 (0.03) & 1.192 (0.02) & 1.536 (0.02) & 0.708 (0.01) \\
		\midrule
% 		\model-NoEdge    & 1.446 (0.03) & 1.165 (0.02) & 1.420 (0.04) & 0.757 (0.02) & 1.666 (0.06) & 1.335 (0.06) & 1.524(0.04) & 0.712(0.02) \\
%         \model-NoSpAttn    & 1.391 (0.01) & 1.122 (0.01) & 1.369 (0.01) & 0.777 (0.00)  & 1.521 (0.03) & 1.222 (0.03)  & 1.478 (0.05)  &	0.732 (0.02) \\
        \B{\model}    &	\B{1.359 (0.03)} & \B{1.093 (0.02)} & \B{1.347 (0.03)} & \B{0.786 (0.01)} &	\B{1.469 (0.01)} &	\B{1.189 (0.02)}  & \B{1.429 (0.03)}  & \B{0.753 (0.01)} \\
		\bottomrule
	\end{tabular}}
		\caption{Experimental results of DTA prediction on PDBbind datasets.
	\vspace{-1mm}
	%Baseline methods are 
	}
	
% 	\scriptsize
	\label{table-expriemnt}
	\vspace{-2mm}
\end{table*}

%% file: section/sec6-conclusion.tex
\section{Conclusion}

In this paper, we propose a novel distance-aware molecule graph attention network (\model) to predict the drug-target binding affinity. We first construct a spatial-enhanced pocket-ligand graph (\graph) to preserve more spatial information and interactions between drug and protein. Moreover, the well-designed \model adopts a hierarchical attention structure, which contains edge-level aggregation and node-level aggregation to capture the unique spatial correlation among atoms. Extensive experimental results on two PDBbind datasets show that \model significantly outperforms all baselines for DTA prediction.